# Evidence of Space weather in Radon Decay

(Unexpected Correlations between GSI $^{222}$Rn Data and Protons Detected by EPAM)

C. Scarlett[a], E. Fischbach [b], B. Freeman[b], R. Olando[b], J. J. Coy[c], P. Edwards[d], R. Burkhart [e], O. Piatibratova[f], T. Monsue[g], D. Osborne[a], J. Edwards[a], L. Mwibanda[a], and A. Alsayegh[a]

[a] Department of Physics, Florida A&M University

[b] Department of Physics and Astronomy, Purdue University

[c] Department of Computer Science, Ball State University

[d] Department of Physics, College of Coastal Georgia

[e] Independent Researcher

[f] Geological Survey of Israel

[g] NASA Goddard Space Flight Center

## ABSTRACT:

The Electron, Proton and Alpha Monitor (EPAM), located at the L1 Position approximately 1-million miles from the earth in the direction of the sun, was designed to detect fluctuations in solar output through counting the numbers of various particles hitting the detector. The EPAM detector is part of an early warning system that can alert the earth to coronal mass ejection events that can damage our electronic grids and satellite equipment. EPAM gives a real-time estimate of changes in the local solar magnetic field directed towards the earth, recorded in the fluctuations of solar particles being ejected. This paper presents an analysis of fluctuations in data taken by the Geological Survey of Israel (GSI) compared to the changes in detected numbers of protons as seen by EPAM. Surprisingly, the GSI and EPAM detectors show an unexpected correlation between the variation in count rate detected by the GSI detectors and an increased numbers of protons seen at EPAM – well above statistical significance of $5\sigma$, indicating a non-random connection between the data sets. The statistically significant overlap between data taken by these two detectors, subject to very different conditions, may hint at a Primakoff mechanism whereby exotic particles, e.g. galactic Dark Matter, couple through magnetic fields to both photons and nuclei. This work builds on an earlier paper on the observations of Radon decay and their implications for particle physics.

**Introduction:**

The Geological Survey of Israel (GSI) developed an experiment to search for anomalies in Radon ($^{222}Rn$) decays over an eleven year period of time. The GSI data reviewed here considers just the gamma rays emitted from daughter particles of the radon; which were detected by NaI detectors. Much of the GSI data has previously been reviewed by both the Lead Scientist, G. Steinitz, and subsequently by Solar Physicist, P. Sturrock et al. The data showed the eleven year solar cycle prominently, indicating a connection between the observed nuclear decay and solar physics. Sturrock postulated the connection may be indicative of solar neutrinos impacting on nuclear decays at the GSI detector on earth. This paper looks at the correlations between excitations and depletions, significant changes, in the GSI measured $^{222}Rn$ count rate as compared to enhanced protons count observed by the Electron Proton Alpha Monitor (EPAM) detector. Significant changes in the GSI count rate, if due to a physics effect, could indicate oscillations in background neutral, weakly interacting particles, e.g. neutrinos, axions or galactic ALPs. While enhancement in the proton count seen by EPAM is associated with Coronal Mass Ejections (CMEs) and other surface solar activity, e.g. solar winds, solar flares and coronal holes. One possible physics connection between these two data sets is the Primakoff production, steering and focusing of neutral weakly interacting matter through the coupling of solar photons to solar magnetic fields. If such an effect is being observed, this could explain observations of the eleven year cycle in the GSI data as first reported by Steinitz and Sturrock.

**GSI Data Analysis:**

To extract the times for which the GSI data showed either a strong excess or depletion in the count rate, the data was first averaged over a two week window that was used to defined "Canonical Day." Figure 1 (Top Left) shows a sample of the GSI data, taken in January of 2008, to illustrate the process. A two week window of data was used to determine how the count rate should vary over the period of a single, 24 hour day. The averaged daily variation per hour is shown in Figure 1 (Top Right). The actual process included: 1. A pedestal is subtracted by using a straight line with endpoints at times $t_i$ = 12:00am and $t_f$ = 11:45pm for each day, 2. The hourly variation above the pedestal is averaged for 14 consecutive days, and 3. The new "Canonical Day" is fitted to data throughout the year using a sinusoidal envelop function as seen in Figure 1 (Bottom). Additionally, a random noise is added to the canonical day each 24 hours to give a better Monte Carlo approach. The data from GSI can be overlaid to determine the degree of fit to the Simulated GSI (GSI Sim) resulting from this approach.

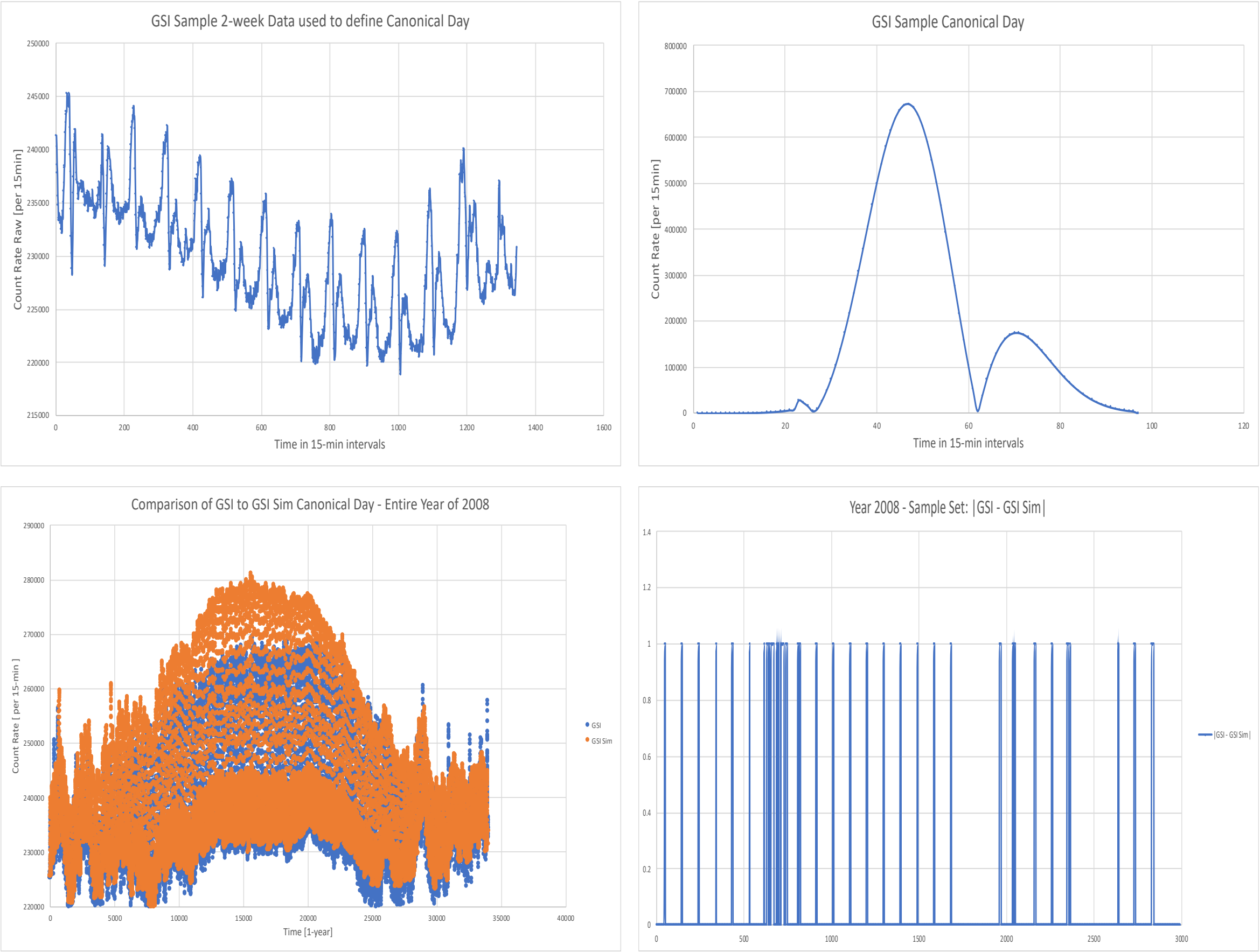

Figure 1: GSI Data from January of 2008 Raw (Top Left), GSI Canonical Day based on a 14-day period as shown, (Top Right), 2008 data for GSI overlaid with GSI Simulated data (Bottom Left) and 2008 Sample Data showing points selected after a cut on the | GSI – GSI Sim | of > 11,000 (counts) has been applied (Bottom Right).

A cut can now be applied to the absolute value of the difference between the expected data set, GSI Sim, and the actual GSI data. This extracts all instances where the GSI showed either an excess or depletion in the count rate compared to the expected values (Canonical Day). A requirement that the GSI data varied from the simulation by at least 11,000 counts per 15-minute period was applied. Whenever the cut was passed, the time period was labeled ‘1’ and if the data was within the cut range the time period was labeled ‘0’ as seen in Figure 1 (Bottom Right).

Ten years of GSI data (2008 – 2017) was processed using the steps illustrated for 2008. Once all data was processed, the extracted time series for GSI could be compared to a time series for EPAM, where a similar analysis had been applied to determine excess proton counts.

**EPAM Data Analysis:**

Similar to GSI, the Electron Proton Alpha Monitor (EPAM) data for protons passing the detector at L1 showed significant fluctuations over the course of a year. Figure 2 shows raw data from 2008.

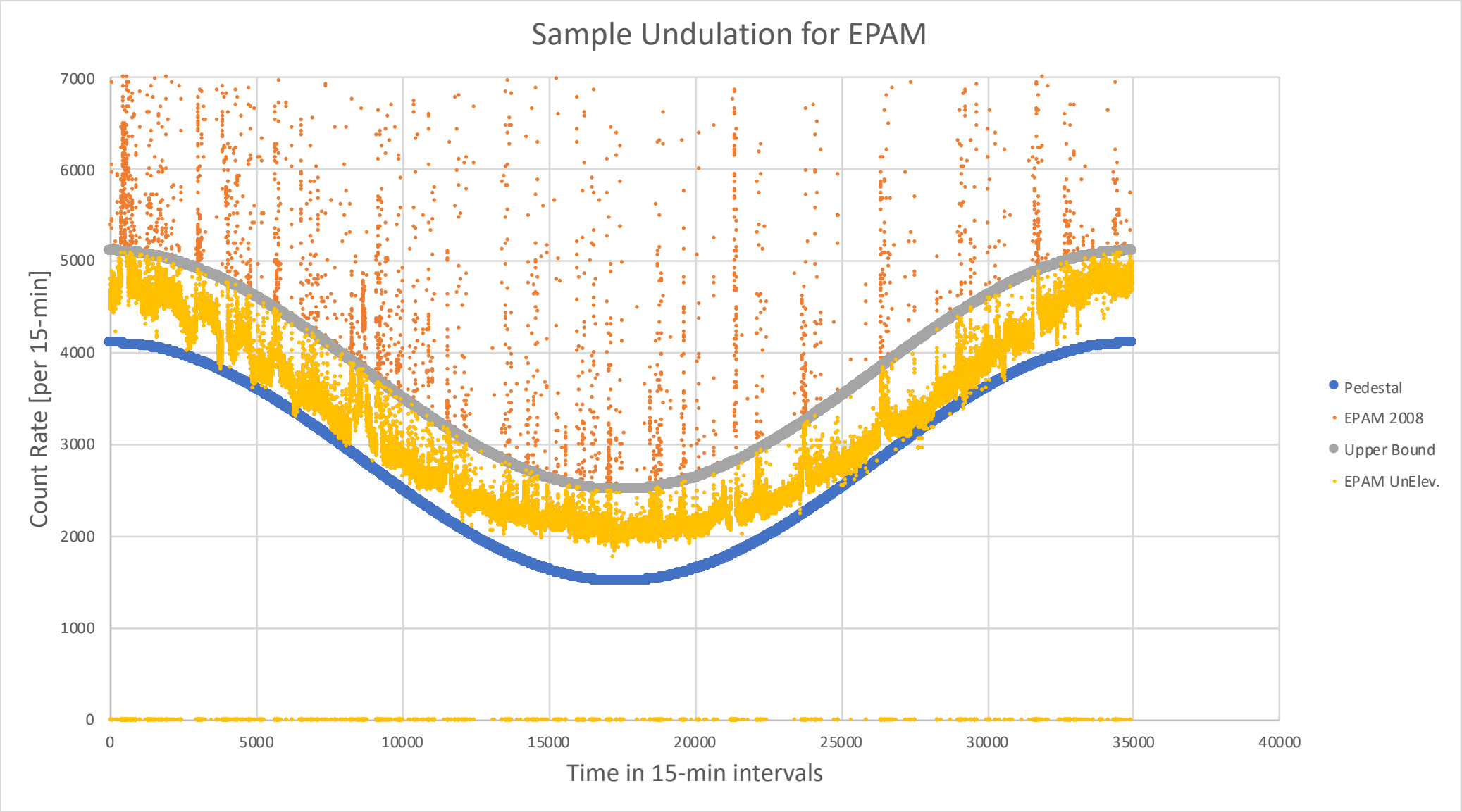

Figure 2: Sample proton count data (orange and gold) from the EPAM detector showing undulation throughout the year. The annual pedestal values (blue) and an applied cut (gray) have been overlaid to show the events identified as expected (gold) and those identified as above threshold (orange).

As with the GSI data, whenever the count rate of protons at the EPAM detector exceeded expectations, Figure 1 (orange), that time period was labeled '1' and whenever the data fell below the cut, Figure 1 (gold), that time period was labeled '0'. This created a time series of '1' and '0' as with the GSI data for comparison.

**Overlap of Time Series Points:**

A comparison of the GSI and EPAM time series appears in Table 1 below. When the two series are unshifted relative to one another, there is a significant overlap between the appearance of elevations in the proton count rate at EPAM and anomalous changes in the gamma count rate in the GSI detector. The algorithm used to determine the degree of overlap between the GSI time series ($G_i$) and the EPAM times series ($E_i$) involves sum of a point-by-point product of the two series, labeled O, and can be written as:

1. $$O = \sum_{i=1}^{N} G_i \times E_i$$

This product is compared to the probability that events labeled '1' in both data sets will randomly appear at the same time $P_r$:

2. $$P_r = \frac{\sum_{i=1}^{N} G_i}{N} \times \frac{\sum_{i=1}^{N} E_i}{N}$$

which when multiplied by the length of the series (N) gives the expected number of overlapping points, denoted as E:

3. $$E = N \times P_r$$

An excess or deficit occurs whenever the actual overlap calculated by 1 is greater or less than what was expected as calculated by 2. To determine the statistical significance of any excess or deficit, one simply has to subtract the value for the expected overlap (E) from what is actually observed (O) and divided by the square-root of what was expected.

4. $$\sigma = \frac{O-E}{\sqrt{E}}$$

The table cites the statistical significance determine in this manner; Figure 3 also displays the overlap excess for a sliding window of time in terms of the statistical significance.

It should be reiterated that these two detectors are spatially almost 1 Million miles apart and that EPAM neither circles the Earth nor is subject to the Earth's atmosphere. These facts, coupled with the lack of observations of similar oscillations in NaI detector data, when performing measurements on other gamma ray sources, makes this correlation quite possibly an indication of a physical process.

| | STATS D |
|---|---|
| P1 = Prob. GSI08 | 0.04712571 |
| P2 = Prob. EPAM09 | 0.24105143 |
| Prod P1*P2 | 0.01135972 |
| Eval = Expect Value | 3975.90226 |
| Actuals | 4323 |
| Sig(Eval - Actual) | 5.50470351 |

Table 1: Statistical comparison between times series defining anomalies in GSI count rate and elevations in the EPAM proton count rate.

Comparison of the GSI and EPAM time series as defined show a statistical significance of 5.5$\sigma$ when the two series start at the same time (Jan. 01, 2008 at 12:00am). Figure 3 shows what happens if the time series windows are shifted relative to one another. Light-mass (sub-milli-eV), weakly interacting particles created in or around the Sun can have ultra-relativistic velocities, moving close to the speed of light. Whereas protons ejected from the Sun during magnetic reconnections, coronal holes or solar flares can have velocities typically from 100-10000 km/s. The lower energy range of protons counted by EPAM has been used in the constructed time series, which has velocities ~ 3150 km/s. These protons can take hours to days to reach the L1 position. Thus, there can be a significant time delay between the arrival of a GSI event versus the arrival of protons at EPAM.

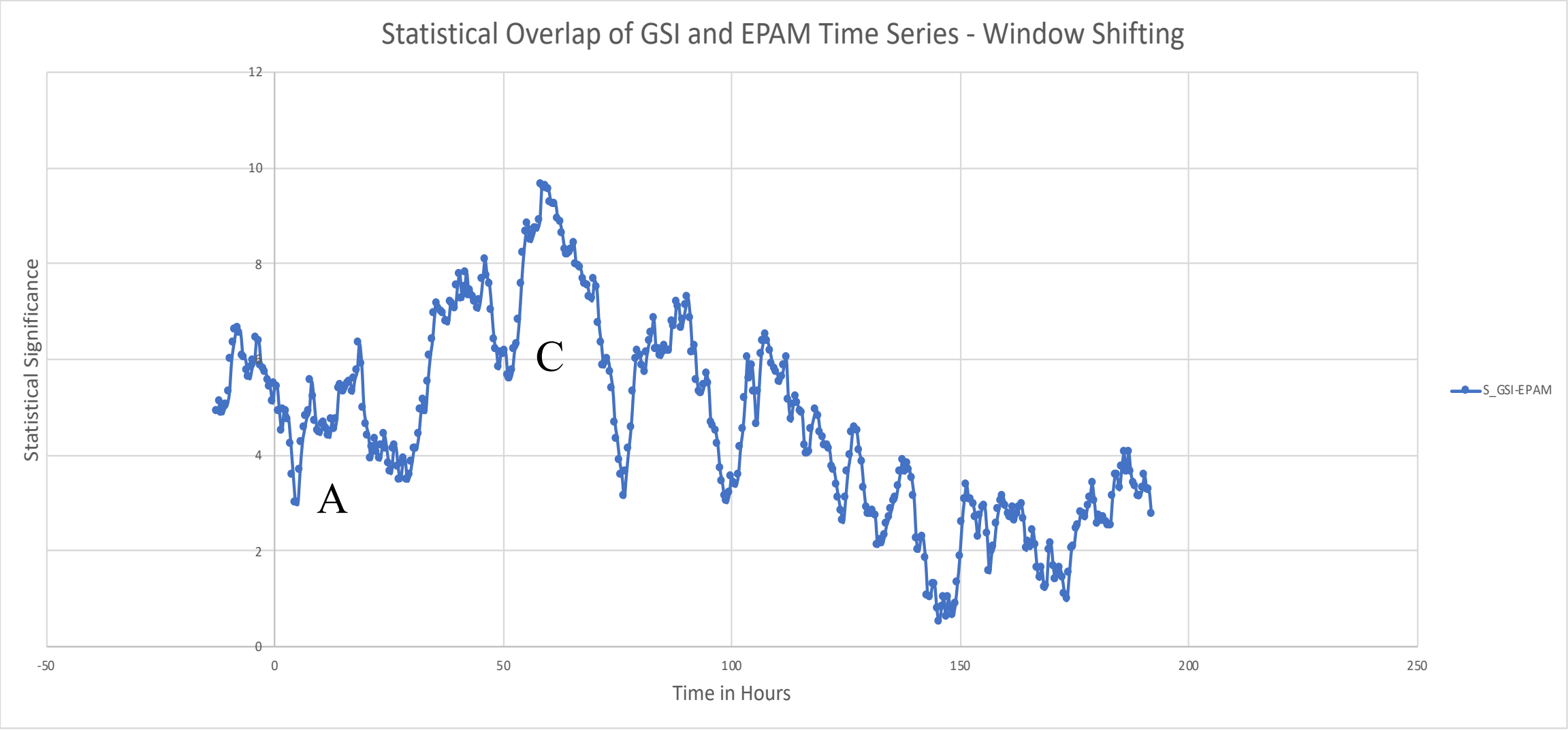

Figure 3: The statistical significance of excessive overlap between the GSI and EPAM time series for a sliding time window. Right of zero, EPAM data is shifted to occur at a later time (x-axis) than the GSI event. Left of zero, GSI data is shifted to occur at a later time than the EPAM event. The observed excess in overlap between these two time series is plotted as a function of its statistical significance (y-axis). Points identified by Peak A can be compared to those in Peak C.

Figure 3 shows that whenever the two time series are shifted relative to one another, the process results in an oscillation in the degree of overlap. A detailed comparison between the GSI-EPAM overlapped data points identified in Peak (A), defined by time values between t=5 and t=20, to those in Peak (C), defined by time values between t=55 and t=70, shows that there is about a 20% difference (~ 850+ distinct time periods) in the total points associated with these two peaks. The identified excess in overlapping points between GSI and EPAM for either peak numbers less than 600. This could indicate that the two peaks come from two distinct dynamical processes and not from some form of oscillation in either time series – further analysis on sliding window techniques for times series where multivariable process occur will be completed.

One possible explanation of these results is the creation of or focusing of weakly interacting particles in the vicinity of the sun. Weakly interacting particles such as neutrinos are regularly detected on Earth using nuclear materials: $^{71}Ga \rightarrow {}^{71}Ge$. If these particles are part of the galactic dust, the natural motion of the Sun could lead to an exchange of energy with this material through the Primakoff mechanism connecting photonic, magnetic and weakly interacting particles. This in turn could explain the imprinting of solar magnetism onto the GSI data set as a whole. Due to the complex interactions that the sun can have with any weakly interacting particles thought to fill the galaxy, e.g. neutrinos, axions and other ALPs, the arrival times for protons at the EPAM detector when compared to particles impacting GSI may show multiple dynamical effects.

**Conclusion:**

The absence of strong oscillations, as observed for radon decay in the GSI data, in other data sets measuring decays of different nuclear isotopes, implies that radon may have properties that make it more sensitive to environmental factors. The GSI data has already been shown to display the 11-year solar cycle (see work of Sturrock et al.) which supports the notion that the data is influenced by solar magnetism. These facts should be followed up on by future experiments to both verify the GSI data in a hermetically sealed system and to search for environmental factors that can directly influence radon decay.

The EPAM detector, which measures protons streaming from the Sun, records data that is heavily impacted by solar magnetic effects. At a position of almost 1 Million miles from Earth, there are few reasons to expect to see any correlations between the EPAM data and any data measuring nuclear decays with instruments on the surface of the Earth. What is most fascinating and unexpected in this analysis, the EPAM proton count rate data shows a strong correlation with the count rate for gammas emitted from a chain decay process of $^{222}Rn$, as seen by the GSI instruments.

We wish to acknowledge the very helpful conversations with Dr. Dennis Krause!